\documentclass[a4paper,11pt,twoside,pdftex]{scrartcl}
\usepackage[title,titletoc]{appendix}
\usepackage{a4wide}
\usepackage{listings}
\usepackage{amsmath}
\usepackage{amsthm}
\usepackage{amssymb}
\usepackage{mathtools}
\usepackage{graphicx}
\usepackage{xcolor}
\usepackage{subfig}
\usepackage[hyperfootnotes=false]{hyperref}
\usepackage{placeins}
\newtheorem{remark}{Remark}
\definecolor{darkblue}{rgb}{0,0,.6}
\definecolor{darkred}{rgb}{.6,0,0}
\definecolor{darkgreen}{rgb}{0,.4,0}
\definecolor{red}{rgb}{.98,0,0}
\definecolor{blue}{rgb}{0.3,0.3,0.9}
\definecolor{violet}{rgb}{0.5,0,0.5}
\newcommand{\vecf}{\boldsymbol{f}}
\newcommand{\vecn}{\boldsymbol{n}}
\newcommand{\vecq}{\boldsymbol{q}}
\newcommand{\vecx}{\boldsymbol{x}}

\newcommand{\vecw}{\boldsymbol{w}}

\newcommand{\vecSigma}{\boldsymbol{\Sigma}}
\newcommand{\dx}{d\boldsymbol{x}}
\newcommand{\ds}{d\boldsymbol{s}}
\newcommand{\vecQ}{\boldsymbol{Q}}
\newcommand{\vecZero}{\boldsymbol{0}}
\newcommand{\matA}{\mathbf{A}}

\newcommand{\matD}{\mathbf{D}}
\newcommand{\matE}{\mathbf{E}}

\newcommand{\matF}{\mathbf{F}}
\newcommand{\matG}{\mathbf{G}}
\newcommand{\matM}{\mathbf{M}}

\newcommand{\numElemOneD}{N_e^{\mathrm{\tiny{1D}}}}
\newcommand{\CG}{\textsf{CG}}
\newcommand{\DG}{\textsf{DG}}
\newcommand{\HDG}{\textsf{HDG}}
\newcommand{\PCG}{\textsf{PCG}}
\title{\Huge Combined \CG{}-\HDG{} Method for Elliptic Problems: Performance Model}
\author{Martin Vymazal$^1$ \and David Moxey$^2$ \and Chris Cantwell$^1$ \and Spencer Sherwin$^1$ \and Robert M. Kirby$^{3}$}
\date{\today}
\date{November 22, 2018}
\begin{document}
\maketitle
%
%
%
\footnote{$^1$Department of Aeronautics, Imperial College London, London, UK}
\footnote{$^2$College of Engineering, Mathematics and Physical Sciences, University of Exeter, UK}
\footnote{$^3$School of Computing, Univ. of Utah, Salt Lake City, UT, USA}
\vspace*{-3em}
\begin{abstract}
We combine continuous and discontinuous Galerkin methods in the setting of a model diffusion problem. 
Starting from a hybrid discontinuous formulation, we replace element interiors by more general subsets 
of the computational domain - groups of elements that support a piecewise-polynomial continuous 
expansion. This step allows us to identify a~new weak formulation of Dirichlet boundary condition in 
the continuous framework. We examine the expected performance of a Galerkin solver that would use 
continuous Galerkin method with weak Dirichlet boundary conditions in each mesh partition and connect 
partitions weakly using trace variable as in \HDG{} method.
\end{abstract}
%
\section{Motivation for combining \CG{} and \HDG{}}
%
%
High-order methods on unstructured grids are now increasingly being used to improve the accuracy of 
flow simulations since they simultaneously provide geometric flexibility and high fidelity. We are 
particularly interested in efficient algorithms for incompressible Navier-Stokes equations that employ 
high-order space discretization and a time splitting scheme. The cost of one step in time is largely 
determined by the amount of work needed to obtain the pressure field, which is defined as a solution 
to a~scalar elliptic problem. Several Galerkin-type methods are available for this task, each of them 
have specific advantages and drawbacks.

\begin{figure}[htpb]
  \centering
  \includegraphics[width=0.95\textwidth]{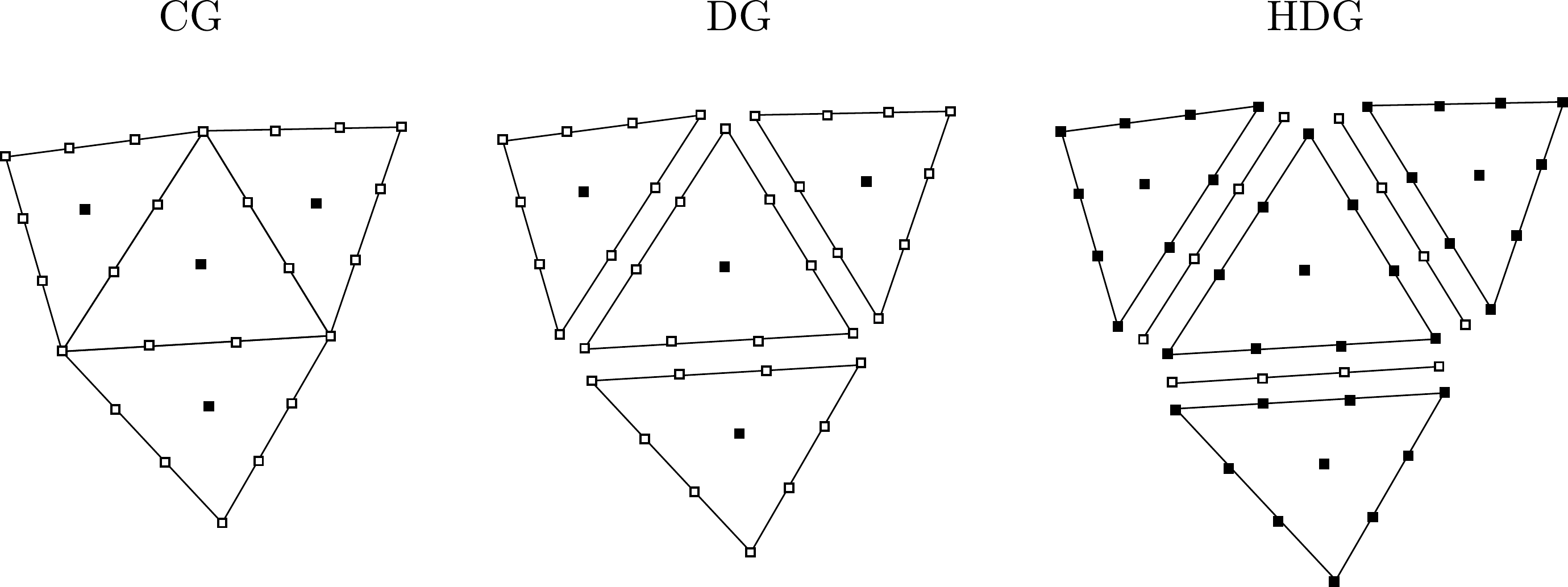}
  \caption{Distribution of unknowns for continuous and discontinuous Galerkin methods.}
  \label{fig_cg_dg_hdg_dofs}
\end{figure}

High-order continuous Galerkin (\CG{}) method is the oldest. Compared to its discontinuous 
counterparts, it involves a smaller number of unknowns (figure \ref{fig_cg_dg_hdg_dofs}), especially 
in a~low-order setting. The \CG{} solution can be accelerated by means of static condensation, which 
produces a globally coupled system involving only those degrees of freedom on the mesh skeleton. The 
element interior unknowns are subsequently obtained from the mesh skeleton data by solving independent 
local problems that do not require any parallel communication.
The amount of information interchanged while constructing and solving the statically condensed system, 
however, is determined by the topology of the underlying grid. Unstructured mesh generators often 
produce meshes with high vertex valency (number of elements incident to given vertex) and CG therefore 
has rather complex communication patterns in parallel runs, which has a negative impact on scaling  
\cite{Yakovlev2016}.

Discontinuous Galerkin (\DG{}) methods \cite{Arnold2002Unified}, on the other hand, duplicate discrete 
variables on element boundaries, thus decoupling mesh elements and requiring at most pairwise 
communication between them. This is at the expense of larger linear system and more time spent in the 
linear solver. Discontinuous discretization is therefore expected to scale better on parallel computers, 
but the improved scaling is not necessarily reflected in significantly smaller CPU times when compared 
to a~\CG{} solver.

Hybrid discontinuous Galerkin (\HDG{}) methods \cite{Cockburn2009} address this problem by introducing 
an additional (hybrid) variable on the mesh skeleton. The hybrid degrees of freedom determine the rank 
of the global system matrix and \HDG{} therefore produces a statically condensed system that is 
similar in size to the \CG{} case. In contrast with \CG{}, the static condensation in \HDG{} takes 
place by construction rather than being an optional iterative technique. Similarly to the classical 
\DG{} method, \HDG{} scales favourably in comparison with \CG{}, but the work-to-communication ratio 
is once again improved due to increased amount of intra-node work rather than due to better overall 
efficiency.

To maximize the potential of each Galerkin variant in a unified setting, we study a~finite element 
discretization that combines the continuous and discontinuous approach by considering a~hybrid 
discontinuous Galerkin method applied to connected groups of elements supporting a globally continuous 
polynomial basis. This settings leads naturally to a~formulation of weak Dirichlet boundary 
conditions for the \CG{} method.

The next section reviews the Hybridizable Discontinuous Galerkin Method which is then modified in 
order to obtain the mixed \CG{}-\HDG{} solver.
%
%
%
%
\section{Overview of the formulation of HDG method}
\label{sec_formulation_hdg_method}
%
We begin with a brief recap of the standard HDG formulation for a~finite element mesh, following 
a~similar approach to that taken in~\cite{Kirby2011} and~\cite{Yakovlev2016}. 

\subsection{Continuous problem}
We seek the solution of a~Poisson equation as a~representative elliptic problem 
\begin{align}
    -\nabla^2 u(\vecx)         & = f(\vecx)   &&  \vecx \in \Omega,
    \label{eq_elliptic_problem_statement}\\
    u(\vecx)                   & = g_D(\vecx) &&  \vecx \in \partial \Omega_D, \nonumber \\
   \vecn \cdot \nabla u(\vecx) & = g_N(\vecx) &&  \vecx \in \partial \Omega_N, \nonumber
\end{align}
on a domain $\Omega$ with Dirichlet ($\partial\Omega_D$) and Neumann ($\partial\Omega_N$) boundary 
conditions, where $\partial \Omega_D \bigcup \partial \Omega_N = \partial \Omega$ and 
$\partial \Omega_D \bigcap \partial \Omega_N = \emptyset$. To formulate the HDG method, we consider 
a~mixed form of \eqref{eq_elliptic_problem_statement} by introducing an auxiliary variable 
$\vecq = \nabla u$:
\begin{align}
  -\nabla \cdot \vecq &= f(\vecx)        && \vecx \in \Omega,
                \label{eq_mixed_form_primal_variable}\\
                \vecq &= \nabla u(\vecx) && \vecx \in \Omega,
                \label{eq_mixed_form_flux_variable}\\
                u(\vecx) &= g_D(\vecx)   && \vecx \in \partial \Omega_D,\\
       \vecq \cdot \vecn &= g_N(\vecx)   && \vecx \in \partial \Omega_N.
\end{align}
The gradient variable $\vecq$ is approximated together with the primal variable $u$, which contrasts 
with the \CG{} method and other discontinuous methods for \eqref{eq_elliptic_problem_statement}.
%
\subsection{HDG interpolation spaces and discretization}
\label{subsec_hdg_interpolation_spaces}
%
%
We limit ourselves to two-dimensional problems for sake of simplicity, but the formal description remains 
unchanged in three dimensions.  
\begin{figure}
  \centering
  \includegraphics[width=0.70\textwidth]{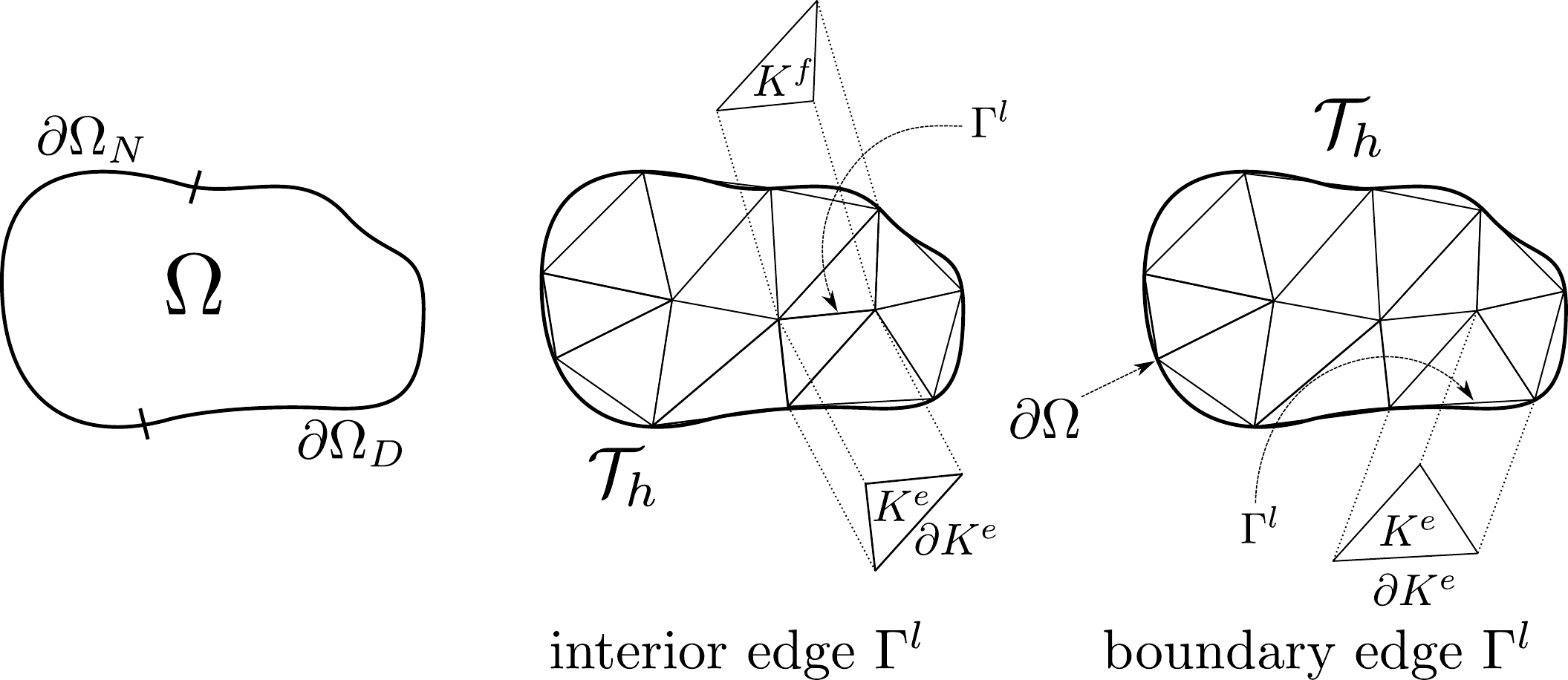}
  \caption{Computational domain and its tesselation demonstrating notation used in the text.}
  \label{fig_domain_tesselation}
\end{figure}
We assume that in the discrete setting, the computational domain $\Omega$ is approximated by its 
tesselation $\mathcal{T}_h$ consisting of non-overlapping and conformal elements $K^e$ such that 
for each pair of distinct indices $e_i \neq e_j$, $K^{e_i} \cap K^{e_j} = \emptyset$. The symbol 
$\Gamma^l$ denotes an interior edge of the tesselation $\mathcal{T}_h$, i.e. an edge 
$\Gamma^l = \bar{K}^i \cap \bar{K}^j$ where $K^i$ and $K^j$ are two distinct elements of the 
tesselation. We say that $\Gamma^l$ is a~boundary edge of the tesselation $\mathcal{T}_h$ if 
there exists an element $K^{e}$ such that $\Gamma^l = K^e \cap \partial \Omega$ and the length of 
$\Gamma^l$ is not zero, as shown in figure~\ref{fig_domain_tesselation}. The set of all internal 
edges is denoted by $\mathcal{E}_h^0$, while $\mathcal{E}_h^{\partial}$ is a~set of all boundary 
edges. Their union $\mathcal{E}_h$ comprises of all mesh edges, 
$\mathcal{E}_h = \mathcal{E}_h^0 \cup \mathcal{E}_h^{\partial}$.

\begin{figure}
  \centering
  \includegraphics[width=0.5\textwidth]{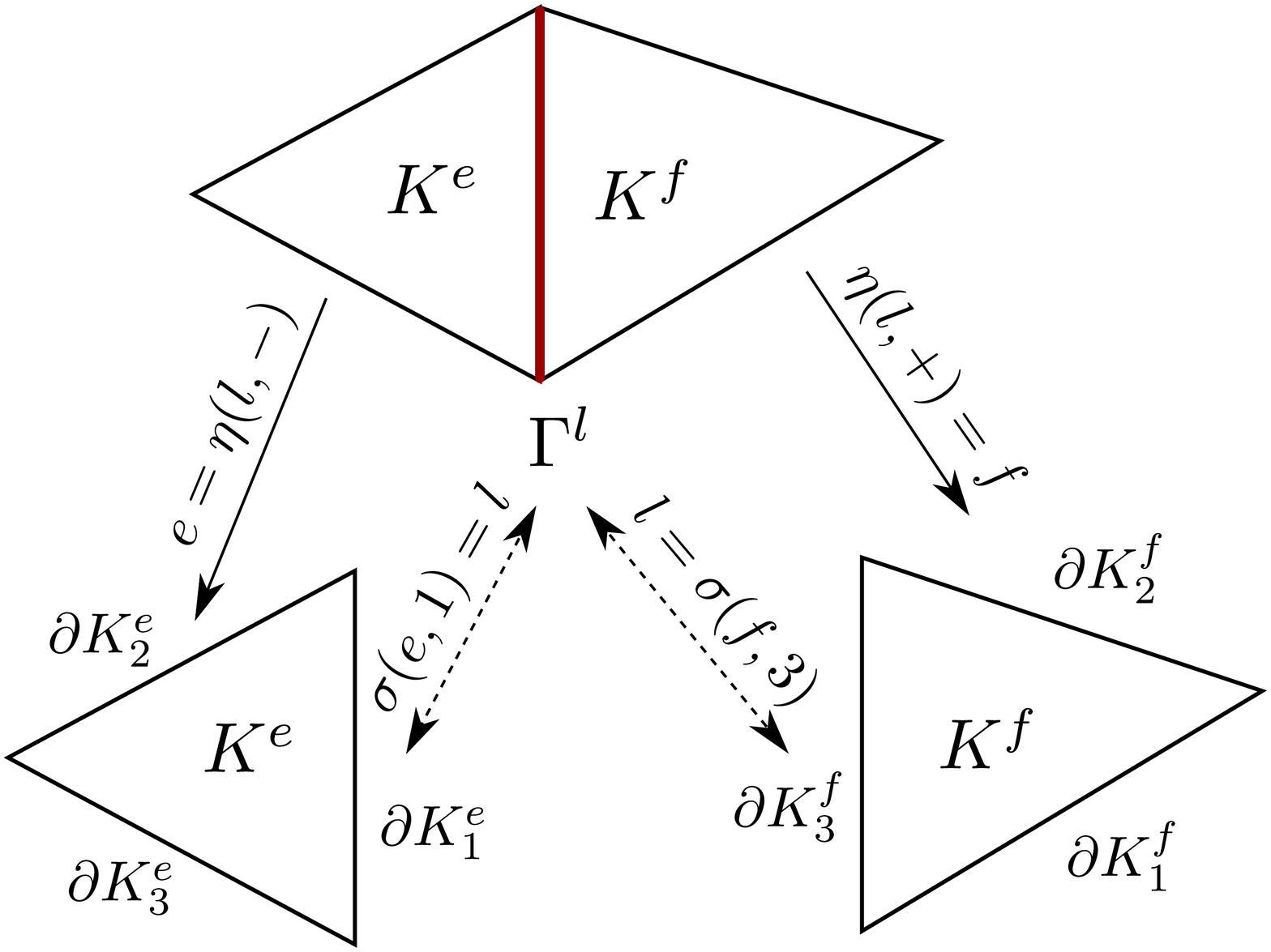}
  \caption{Index mappings relating edge and element ids.}
  \label{fig_index_mappings}
\end{figure}

In order to describe some terms in the \HDG{} formulation, it is also useful to introduce 
mappings that relate elements to their local edges, as shown in figure~\ref{fig_index_mappings}. 
Let $\partial K_j^e$ be the $j$-th edge of element $K^e$, and suppose that this is also the 
$l$-th edge $\Gamma^l$ in the global edge numbering. Then we define the local-to-global edge mapping 
$\sigma$ by setting $\sigma(e,j) = l$ so that we can write $\partial K_j^e = \Gamma^{\sigma(e,j)}$. 
An interior edge $\Gamma^l$ is the intersection of the boundaries of two elements $K^e$ and $K^f$, 
hence we set $\eta(l,+) = e$ and $\eta(l,-) = f$ in order to be able to write 
$\Gamma^l = \partial K^{\eta(l,+)} \cap \partial K^{\eta(l,-)}$.

%
%
\subsection{Approximation spaces}
%
The finite element spaces supported by the (two-dimensional) tesselation $\mathcal{T}_h$ are defined 
as follows:
\begin{alignat*}{4}
  V_h & \coloneqq \{ v \in L^2(\Omega) &&: \quad && v|_{K^e} \in \mathcal{P}(K^e) && \quad \forall K^e \in \mathcal{T}_h \},\\
  \vecSigma_h & \coloneqq \{ \vecw \in [L^2(\Omega)]^2 &&: \quad && \vecw|_{K^e} \in \vecSigma(K^e) && \quad \forall K^e \in \mathcal{T}_h \},\\
  \mathcal{M}_h & \coloneqq \{ \mu \in L^2(\Omega) &&: \quad && \mu|_{\Gamma^l} \in \mathcal{P}(\Gamma^l) && \quad \forall\, \Gamma^l \in \Gamma \},
\end{alignat*}
where $\mathcal{P}(\Gamma^l) = \mathcal{S}_P(\Gamma^l)$ is the polynomial space over the standard segment
\begin{equation*}
 \mathcal{S}_P(\Gamma^l) = \{s^p\ :\  0 \leq p \leq P,\ [x_1(s), x_2(s)] \in \Gamma^l,\ -1 \leq s \leq 1\},
\end{equation*}
and $\mathcal{P}(K^e)$ is the space of polynomials of degree $P$ defined on a~standard region, 
which can either be the standard triangle
\begin{equation*}
  \mathcal{P}(K^e) =  \mathcal{T}_P(K^e) = \{\xi_1^p \xi_w^q\ :\  0 \leq p + q \leq P,\ [x_1(\xi_1, \xi_2), x_2(\xi_1, \xi_2)] \in K^e,\ -1 \leq \xi_1 + \xi_2 \leq 0\},
\end{equation*}
or standard quadrilateral
\begin{equation*}
  \mathcal{P}(K^e) = \mathcal{Q}_P(K^e) = \{\xi_1^p \xi_2^q\ :\ 0 \leq p,q\leq P,\ [x_1(\xi_1, \xi_2), (x_2(\xi_1, \xi_2)] \in K^e,\ -1 \leq \xi_1,\xi_2 \leq 1\}.
\end{equation*}
Similarly $\vecSigma(K^e) = [\mathcal{T}_P(K^e)]^2$ or $\vecSigma(K^e) = [\mathcal{Q}_P(K^e)]^2$. 
There is no requirement on global continuity of the expansion. This is also true for the trace 
space $\mathcal{M}_h$: a discrete variable $\lambda \in \mathcal{M}_h$ is multi-valued at every mesh 
vertex shared by multiple interior edges.
%
%
\subsection{Global formulation for HDG problem}
\label{subsec_discrete_global_solver}
%
%
%
Given an element $K \in \mathcal{T}_h$ and two functions $u,v \in L^2(\mathcal{T}_h)$, we
define their $L^2$ scalar product by
\begin{equation*}
  (u,v)_{\mathcal{T}_h} = \sum\limits_{K\in \mathcal{T}_h} (u,v)_K, \quad \text{ where } \quad 
  (u,v)_K = \int\limits_K uv \,\dx.
\end{equation*}
Similarly, the $L^2$ product of functions $u$ and $v$ that are square-integrable on element traces
are defined by:
\begin{equation*}
 \langle u,v \rangle _{\partial \mathcal{T}_h} 
   = \sum\limits_{K\in \mathcal{T}_h} \langle u, v \rangle_{\partial K}
     \quad \text{ where } \quad \langle u, v\rangle_{\partial K} = \int\limits_{\partial K} uv \,\ds
\end{equation*}

The DG method seeks an approximation pair $(u^{DG},\vecq^{DG})$ to $u$ and $\vecq$, respectively, in 
the space $V_h \times \vecSigma_h$. The solution is required to satisfy the weak form of 
\eqref{eq_mixed_form_primal_variable} and \eqref{eq_mixed_form_flux_variable}
\begin{align}
  \left(\vecq^{DG}, \nabla v\right)_{\mathcal{T}_h} 
    &= \left(f,v\right)_{\mathcal{T}_h} + \left \langle \vecn^e \cdot \tilde{\vecq}^{DG}, v \right \rangle_{\partial \mathcal{T}_h} \\
  \left(\vecq^{DG}, \vecw\right)_{\mathcal{T}_h} &= - \left( u^{DG},\nabla \cdot \vecw \right)_{\mathcal{T}_h} + 
   \left\langle \tilde{u}^{DG},\vecw \cdot \vecn^e \right\rangle_{\partial \mathcal{T}_h}
\end{align}
for all $(v,\vecw) \in V_h(\Omega) \times \vecSigma_h(\Omega)$, where the numerical traces 
$\tilde{u}^{DG}$ and $\tilde{\vecq}^{DG}$ have to be suitably defined in terms of the approximate 
solution $(u^{DG}, \vecq^{DG})$. For details, we refer the reader to~\cite{Cockburn2009}. This 
choice of trace variables allows us to construct the discrete \HDG{} system involving only trace 
degrees of freedom $\tilde{u}^{DG}$. Once $\tilde{u}$ is known, the element-interior degrees of 
freedom represented by both the primal variable $u$ and gradient $\vecq$ can be reconstructed from 
element-boundary values. 

We note that the element-interior variable $u$ restricted to element traces is not equal to the 
hybrid variable $\tilde{u}$, but only approximates it: due to the definition of approximation spaces 
$V_h$ and $\mathcal{M}_h$, $u$ must be continuous along element boundaries, while $\tilde{u}^{DG}$ 
is allowed to have jumps in element vertices. 
%
%
\subsection{Local solvers in the HDG method}
\label{subsec_local_solver}
%
%
Assume that the function
\begin{equation}
  \lambda \coloneqq \tilde{u}^{DG} \in \mathcal{M}_h,
\end{equation}
is given. Then the solution restricted to element $K^e$ is a~function 
$u^e, \vecq^e$ in $P(K^e) \times \vecSigma(K^e)$ that satisfies the following equations:
\begin{align}
  \left(\vecq^e, \nabla v\right)_{K^e} &= (f,v)_{K^e} + \left\langle  \vecn^e \cdot \tilde{\vecq}^e, v \right\rangle_{\partial K^e}
  \label{eq_local_formulation_u}\\
  \left( \vecq^e, \vecw \right)_{K^e}  &= - \left( u^e, \nabla \cdot \vecw \right)_{K^e} 
  + \left\langle \lambda, \vecw \cdot \vecn^e \right\rangle_{\partial K^e},
  \label{eq_local_formulation_flux}
\end{align}
for all $(v,\vecw) \in P(K^e) \times \vecSigma (K^e)$. For a~unique solution of the above equations to exist, the numerical trace
of the flux must depend only on $\lambda$ and on $(u^e,\vecq^e)$:
\begin{equation}
  \tilde{\vecq}^e(\vecx) = \vecq^e(\vecx) - \tau \bigl(u^e(\vecx) - \lambda(\vecx)\bigr)\vecn^e \quad \text{on  } \partial K^e
  \label{eq_flux_trace}
\end{equation}
for some positive function $\tau$. The analysis presented in \cite{Cockburn2009} reveals that 
as long as $\tau > 0$, its value can be arbitrary without degrading the robustness of the solver. 
For the limiting value of $\tau \rightarrow \infty$, one obtains a~statically condensed continuous 
Galerkin formulation. In this sense, $\tau$ plays the role of a~method selector as opposed to 
traditional penalty parameter used in Nitsche's method, for example.
%
%
\subsection{Global problem for trace variable}
\label{subsec_global_solver_lambda}
%
%
We denote by $(U_{\lambda}, \vecQ_{\lambda})$ and by $(U_f, \vecQ_f)$ the solution to the local problem
\eqref{eq_local_formulation_u}, \eqref{eq_local_formulation_flux} when $\lambda = 0$ and $f = 0$,
respectively. Due to the linearity of the original problem \eqref{eq_elliptic_problem_statement} and 
its mixed form, the solution satisfies
\begin{equation}
  (u^{HDG}, \vecq^{HDG}) = (U_{\lambda}, \vecQ_{\lambda}) + (U_f, \vecQ_f).
\end{equation}
In order to uniquely determine $\lambda$, we require that the boundary conditions be weakly 
satisfied and the normal component of the numerical trace of the flux $\tilde{\vecq}$ given by 
\eqref{eq_flux_trace} is single valued, rendering the numerical trace conservative.

We say that $\lambda$ is the element of $\mathcal{M}_h$ such that
\begin{align}
  \lambda &= \mathbb{P}_h(g_D) \quad \text{on } \partial \Omega_D\\
  \left\langle \mu , \tilde{\vecq} \cdot \vecn \right\rangle_{\partial \mathcal{T}} 
                  &= \left\langle \mu, g_N \right\rangle_{\partial \Omega_N},
  \label{eq_transmission_condition}
\end{align}
for all $\mu \in \mathcal{M}_h^0$ such that $\mu = 0$ on $\partial \Omega_D$. Here $\mathbb{P}_h$ denotes 
the $L^2$-projection into the space of restrictions to $\partial \Omega_D$ of functions of 
$\mathcal{M}_h$. 

In the following, we consider $u^e(\vecx), \vecq^e(\vecx) = [q_1, q_2]^T$ and $\lambda^l(\vecx)$ to 
be finite expansions in terms of basis functions $\phi_j^e(\vecx)$ for the expansions over elements 
and the basis $\psi_j^l(\vecx)$ over the traces of the form:
\begin{equation*}
  u^e(\vecx) = \sum\limits_{j=1}^{N_u^e} \phi_j^e (\vecx) \hat{\underline{u}}^e[j] \qquad
  \vecq_k^e(\vecx) = \sum\limits_{j=1}^{N_q^e} \phi_j^e(\vecx) \hat{\underline{q}}_k^e[j] \qquad
  \lambda^l(\vecx) = \sum\limits_{j=1}^{N_{\lambda}^l} \psi_j^l(\vecx) \hat{\underline{\lambda}}^l[j]
\end{equation*}
%
%
\section{Discrete form of HDG local solver}
\label{sec_discrete_local_solver}
%
%
We now define several local matrices stemming from standard Galerkin formulation, where scalar test 
functions $v^e$ are represented by $\phi_i^e(\vecx)$, with $i=1,\ldots,N_u^e$.
%
%
\begin{align*}
  \matD_k^e[i,j] &= \left(\phi_i^e, \frac{\partial \phi_j^e}{\partial x_k}\right)_{K_e} &
  \matM^e[i,j] &= \left(\phi_i^e, \phi_j^e\right)_{K_e}\\
  \matE_l^e[i,j] &= \left\langle \phi_i^e, \phi_j^e \right\rangle_{\partial K_l^e} &
  \widetilde{\matE}_{kl}^e[i,j] &= \left\langle \phi_i^e , \phi_j^e n_k^e \right\rangle_{\partial K_l^e}\\
  \matF_l^e[i,j] &= \left\langle \phi_i^e, \psi_j^{\sigma(e,l)}\right\rangle_{\partial K_l^e} & 
  \widetilde{\matF}^e_{kl}[i,j] &= \left\langle \phi_i^e, \psi_j^{\sigma(e,l)} n_k^e\right\rangle_{\partial K_l^e}
\end{align*}
If the trace expansion matches the expansions used along the edge of the elemental expansion and 
the local coordinates are aligned, that is $\psi_i^{\sigma(e,l)}(s) = \phi_{k(i)}(s)$ then 
$\matE_l^e$ contains the same entries as $\matF_l^e$ and similarly $\widetilde{\matE}_{kl}^e$ 
contains the same entries as $\widetilde{\matF}_{kl}^e$.

Inserting the finite expansions of the trial functions into equations \eqref{eq_local_formulation_u} 
and \eqref{eq_local_formulation_flux}, and using the definition of the flux \eqref{eq_flux_trace} 
yields the matrix form of \emph{local solvers}
\begin{align}
  \renewcommand*{\arraystretch}{1.5}
  \sum\limits_{k=1,2}
  \left\{ (\matD_k^e)^T - \sum\limits_{l=1}^{N_b^e} \left[ \widetilde{\matE}_{kl}^e \right] \right\}
  \underline{\hat{q}}_k^e
  &+ \sum\limits_{l=1}^{N_b^e} \tau^{e,l} \left[ \matE_l^e \underline{\hat{u}}^e - 
                                                \matF_l^e \underline{\hat{\lambda}}^{\sigma(e,l)}\right] = \underline{f}^e
  \label{eq_HDG_local_solver_a}\\
  \matM^e \underline{\hat{q}}_k^e = -(\matD_k^e)^T \underline{\hat{u}}^e 
                                  &+ \sum\limits_{l=1}^{N_b^e} \widetilde{\matF}_{kl}^e \underline{\hat{\lambda}}^{\sigma(e,l)}
  \qquad k = 1,2
  \label{eq_HDG_local_solver_b}
\end{align}
The \emph{global equation for} $\lambda$ can be obtained by discretizing the transmission condition 
\eqref{eq_transmission_condition}. We introduce local element-based and edge-based matrices
\begin{equation*}
  \overline{\matF}^{l,e}[i,j] = \left\langle \psi_i^l, \phi_j^e\right\rangle_{\Gamma^l} \qquad
  \overset{\simeq}{\matF}^{l,e}_k [i,j] = \left\langle \psi_i^l, \phi_j^e n_k^e\right\rangle_{\Gamma^l} \qquad
  \bar{\matG}^l[i,j] = \left\langle \psi_i^l, \psi_j^l \right\rangle_{\Gamma^l}
\end{equation*}
and define
\begin{equation*}
  \underline{g}_N^l[i] = \left\langle g_n, \psi_i^l\right\rangle_{\Gamma^l \bigcap \partial \Omega_N}.
\end{equation*}
The transmission condition in matrix form is then
\begin{equation*}
  \renewcommand*{\arraystretch}{1.5}
  \left[ \overset{\simeq}{\matF}_1^{l,e} \overset{\simeq}{\matF}_2^{l,e} \right]
  \left[ \begin{array}{c} \underline{\hat{q}}_1^e \\ \underline{\hat{q}}_2^e \end{array}\right] + 
  \left[ \overset{\simeq}{\matF}_1^{l,f} \overset{\simeq}{\matF}_2^{l,f} \right]
  \left[ \begin{array}{c} \underline{\hat{q}}_1^f \\ \underline{\hat{q}}_2^f \end{array}\right] +
  (\tau^{e,i} + \tau^{f,j})\bar{\matG}^l \underline{\hat{\lambda}}^l - 
  \tau^{e,i} \bar{\matF}^{l,e} \underline{u}^e - \tau^{f,j} \bar{\matF}^{l,f} \underline{u}^f = 
  \underline{g}_N^l,
\end{equation*}
where we are assuming that $l = \sigma(e,i) = \sigma(f,j)$.
%
%
\section{Combined Continuous-Discontinuous Formulation}
%
%
To take advantage of the efficiency and lower memory requirements of continuous Galerkin method 
together with the flexibility and more favorable communication patterns of discontinuous Galerkin 
methods in domain-decomposition setting, we combine both as follows. Each mesh partition is seen as 
a~'macro-element', where the governing equation is discretized by continuous Galerkin solver, while 
the patches are coupled together weakly as in HDG. This means that the scalar flux (hybrid variable) 
$\lambda$ is only defined on inter-partition boundaries.
\begin{figure}[htpb]
  \centering
  \includegraphics[width=0.8\textwidth]{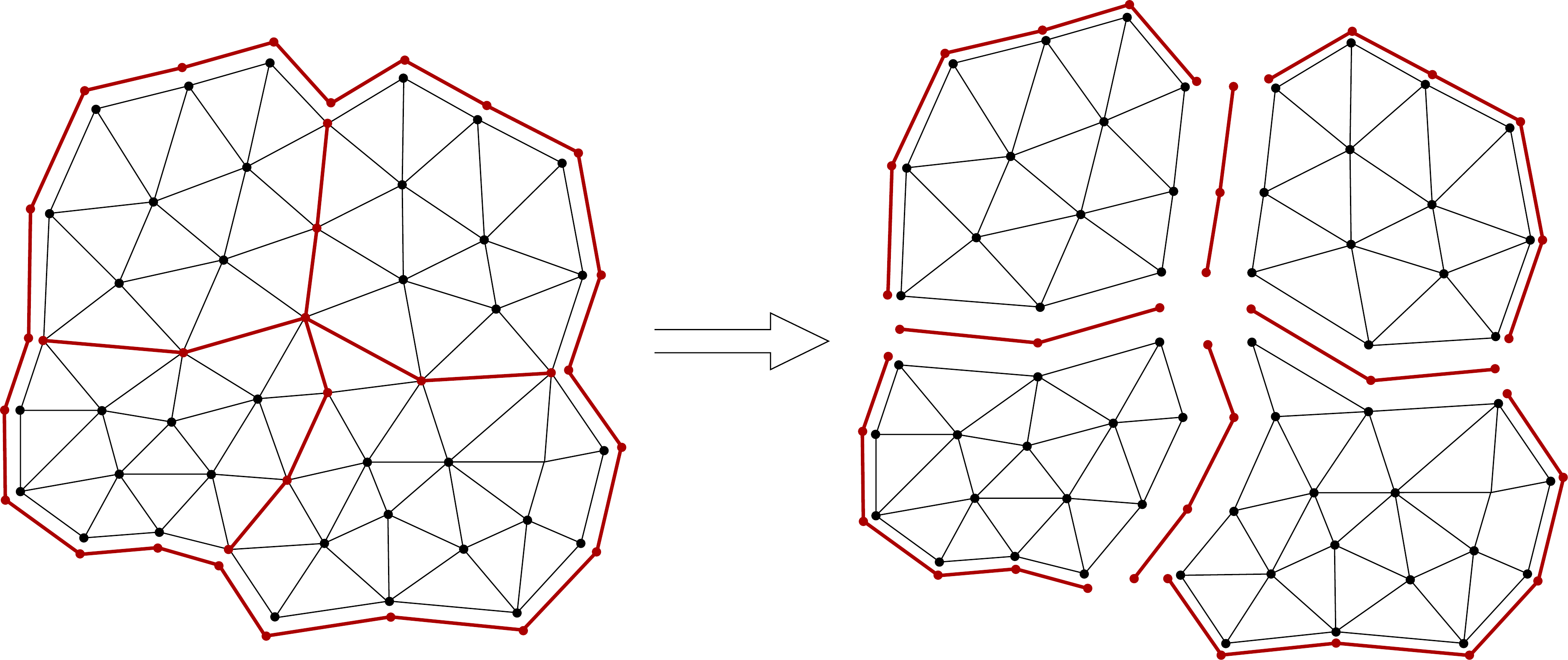}
\end{figure}
%
\section{Continuous-Discontinuous Solver}
%
%
Given the hybrid \CG{}-\HDG{} setup, the \HDG{} local solver defined previously for one element would 
now be applied to a~group of elements supporting a~piecewise-continuous basis. The motivation of 
this section is to take the matrix form of the HDG solver and apply it in such piecewise-continuous 
setting. We will show that the discrete weak form reduces to the 'standard' Laplace operator plus 
extra terms, which will be only applied on elements adjacent to partition boundaries, providing weak 
coupling between each partition and the global trace variable.

We start again from the weak mixed problem \eqref{eq_local_formulation_u},
\eqref{eq_local_formulation_flux}, but integrate the second term in the flux equation
\eqref{eq_local_formulation_flux} by parts once again. This modified flux form allows for a~symmetric
boundary contribution to the linear system as will be explained shortly. In order to distinguish
between the standard \HDG{} local solver within a~single element and \HDG{} applied to the whole
domain tesselation $\mathcal{T}_h$, the superscript `$e$' has been replaced by $\mathcal{T}_h$ where
appropriate. The 'macro element' form yields a~system
\begin{align}
  \left( \vecq^{\mathcal{T}_h}, \nabla v\right)_{\mathcal{T}_h} &= \left(f,v\right)_{\mathcal{T}_h}
    + \left\langle \vecn^{\mathcal{T}_h} \cdot \tilde{\vecq}^{\mathcal{T}_h} , v\right\rangle_{\partial \mathcal{T}_n}
  \label{eq_local_cg_dg_formulation_u}\\
  \left(\vecq^{\mathcal{T}_h}, \vecw \right)_{\mathcal{T}_h} &= \left( \nabla u^{\mathcal{T}_h} , \vecw \right)_{\mathcal{T}_h}
  - \left\langle u^{\mathcal{T}_h} , \vecw \cdot \vecn^{\mathcal{T}_h} \right\rangle_{\partial \mathcal{T}_h}
  + \left\langle \lambda, \vecw \cdot \vecn^{\mathcal{T}_h} \right\rangle_{\partial \mathcal{T}_h}
  \label{eq_local_cg_dg_formulation_flux}
\end{align}
The numerical approximation $u^{\mathcal{T}_h}$ belongs to the space $V_h^{\mathcal{T}_h}$ and
$\vecq^{\mathcal{T}_h}$ lies in $\vecSigma_h^{\mathcal{T}_h}$, which are defined as
\begin{alignat*}{4}
  V_h^{\mathcal{T}} & \coloneqq \{ v \in \mathcal{C}^0(\Omega) &&: &&\quad v|_{K^e} \in P(K^e) &&\quad \forall K^e \in \mathcal{T}_h \},\\
  \vecSigma_h^{\mathcal{T}} & \coloneqq \{ \vecw \in [L^2(\Omega)]^2 &&: && \quad \vecw|_{K^e} \in \vecSigma(K^e) &&\quad \forall K^e \in \mathcal{T}_h \}.
\end{alignat*}
Using the definition of the trace flux
\begin{equation*}
 \tilde{\vecq}^{\mathcal{T}_h}(\vecx) = \vecq^{\mathcal{T}_h}(\vecx) - \tau (u^{\mathcal{T}_h}(\vecx) - \lambda(\vecx)) \vecn^{\mathcal{T}_h} \quad \text{ on } \partial \mathcal{T}_h,
\end{equation*}
and the fact that the integral over $\mathcal{T}_h$ can be written as a~sum of integrals over all 
$K^e \in \mathcal{T}_h$, equations \eqref{eq_local_cg_dg_formulation_u} and
\eqref{eq_local_cg_dg_formulation_flux} become
\begin{align}
  &\sum\limits_{K^e \in \mathcal{T}_h} \,\left(\nabla v, \vecq^e \right)_{K^e}
  - \sum\limits_{\substack{K^e \\ \partial K^e \cap \partial \mathcal{T}_{h,D} \neq \emptyset}}
                                                    \left\langle v, \vecn^e \cdot \vecq^e \right\rangle_{\partial K^e}
  + \tau \mkern-9mu \sum\limits_{\substack{K^e \\ \partial K^e \cap \partial \mathcal{T}_{h,D} \neq \emptyset}}
                                                  \left\langle v,u^e \right\rangle_{\partial K^e} \nonumber\\
  -& \tau \mkern-9mu \sum\limits_{\substack{K^e \\ \partial K^e \cap \partial \mathcal{T}_{h,D} \neq \emptyset}} \:
                                                   \left\langle v,\lambda \right\rangle_{\partial K^e}
  = \sum\limits_{K^e \in \mathcal{T}_h} \left( v,f \right)_{K^e}
  \label{eq_local_cg_dg_formulation_u_elem_sum}\\
  & \sum\limits_{K^e \in \mathcal{T}_h} \left(\vecw, \vecq^e \right)_{K^e}
  + \sum\limits_{\substack{K^e \\ \partial K^e \cap \partial \mathcal{T}_{h,D} \neq \emptyset}}
  \left\langle u^e, \vecw \cdot \vecn^e \right\rangle_{\partial K^e} \nonumber\\
  -& \sum\limits_{K^e \in \mathcal{T}_h} \left( \vecw, \nabla u^e\right)_{K^e}
  - \sum\limits_{\substack{K^e \\ \partial K^e \cap \partial \mathcal{T}_{h,D} \neq \emptyset}} \:
                        \left\langle \vecw \cdot \vecn^e, \lambda \right\rangle_{\partial K^e} = 0
  \label{eq_local_cg_dg_formulation_q_elem_sum}
\end{align}
A~continuous Galerkin solver with Dirichlet data prescribed by the variable $\lambda$ can be
obtained by eliminating the flux variable from the system and reverting back to primal form for the
unknown $u$. The mass matrix which appears in the second equation of the local solver after
evaluating the dot product $\left(\vecq^{\mathcal{T}_h}, \vecw\right)_{\mathcal{T}_h}$ is now
block-diagonal as a~consequence of the discontinuous nature of the discrete flux
$\vecq^{\mathcal{T}_h}$, hence the elimination of $\vecq^{\mathcal{T}_h}$ from the system can be
performed element-wise.  The matrix equivalent of \eqref{eq_local_cg_dg_formulation_u_elem_sum},
\eqref{eq_local_cg_dg_formulation_q_elem_sum} written for a~single element $K^e \in \mathcal{T}_h$
adjacent to Dirichlet boundary reads
\begin{align}
  \renewcommand*{\arraystretch}{1.5}
  \sum\limits_{k=1,2}\left\{
     (\matD_k^e)^T - \sum\limits_{l=1}^{N_b^e} \widetilde{\matE}_{kl}^e
  \right\} \underline{\hat{q}}_k^e
  + \sum\limits_{l=1}^{N_b^e} \tau^{e,l} \left[ \matE_l^e \underline{\hat{u}}^e -
                                                \matF_l^e \underline{\hat{\lambda}}^{\sigma(e,l)}\right] = \underline{f}^e
  \label{eq_local_cg_dg_formulation_u_matrix}\\
  \matM^e \underline{\hat{q}}_k^e = \left\{ (\matD_k^e) - \sum\limits_{l=1}^{N_b^e} \widetilde{\matE}_{kl}^e  \right\}
                                    \underline{\hat{u}}^e
                                    + \sum\limits_{l=1}^{N_b^e} \widetilde{\matF}_{kl}^e \underline{\hat{\lambda}}^{\sigma(e,l)}
  \qquad k = 1,2
  \label{eq_local_cg_dg_formulation_q_matrix}
\end{align}
The discrete flux $\underline{\hat{q}}_k^e$ expressed from \eqref{eq_local_cg_dg_formulation_q_matrix}
and substituted in equation $\eqref{eq_local_cg_dg_formulation_u_matrix}$ yields element-wise
contribution to the left- and right-hand side of the linear system which can be expressed as
\begin{align*}
 &\sum\limits_{k=1,2} \Biggl\{ \underbrace{(\matD_k^e)^T \bigl(\matM^e\bigr)^{-1} \matD_k^e}_{\boxed{1}}
  %
          -   \underbrace{
                \biggl(\sum\limits_{l=1}^{N_b^e} \tilde{\matE}_{kl}^e\biggr)
                \bigl(\matM^e\bigr)^{-1}
                \matD_k^e
              }_{\boxed{2a}}
           \Biggr\}\underline{\hat{u}}^{e}\\
  %
 -&\sum\limits_{k=1,2} \Biggl\{
              \underbrace{
                          (\matD_k^e)^T \bigl(\matM^e\bigr)^{-1} \biggl(\sum\limits_{l=1}^{N_b^e} \tilde{\matE}_{kl}^e\biggr)
                        }_{\boxed{2b}}
  %
  %
            + \underbrace{
                          \biggl(\sum\limits_{l=1}^{N_b^e} \tilde{\matE}_{kl}^e \biggr) \bigl(\matM^e\bigr)^{-1}
                          \biggl(\sum\limits_{l=1}^{N_b^e} \tilde{\matE}_{kl}^e \biggr)
                       }_{\boxed{3}}
               \Biggr\} \underline{\hat{u}}^{e}
              + \underbrace{
                        \sum\limits_{l=1}^{N_b^e} \tau^{(e,l)}\matE_l^e \underline{\hat{u}}^e
                }_{\boxed{4}}
  \\
  =& \underline{\vecf}^e
      + \sum\limits_{k=1,2}
        \left\{
          \biggl(\sum\limits_{l=1}^{N_b^e} \tilde{\matE}_{kl}^e - (\matD_k^e)^T \biggr)
          \bigl(\matM^e\bigr)^{-1}
          \biggl(\sum\limits_{l=1}^{N_b^e} \tilde{\matF}_{kl}^e\underline{\hat{\lambda}}^{\sigma(e,l)}\biggr)\right\}
      + \sum\limits_{l=1}^{N_b^e} \tau^{(e,l)}\matF_l^e \underline{\hat{\lambda}}^{\sigma(e,l)}
\end{align*}
Term $\boxed{1}$ on the left-hand side is a~discrete Laplacian that arises from the standard continuous
Galerkin discretization, which would typically be accompanied by the forcing term $\underline{\vecf}^e$
on the right hand side. The additional numbered terms represent a~\emph{weak imposition of Dirichlet boundary 
data represented by} $\underline{\hat{\lambda}}$. (More details on weak Dirichlet boundary conditions can be 
found in our paper \cite{VymazalWeakBC}). This new expression therefore denotes a modification of the existing 
matrix system and right hand side, which makes implementation relatively straightforward. The matrix
expressions $\boxed{2a}$, $\boxed{2b}$, $\boxed{3}$ and $\boxed{4}$ appear in the formulation only
for elements $K^e$ containing at least one edge on Dirichlet boundary of $\Omega$. In addition,
expressions $\boxed{3}$ and $\boxed{4}$ are symmetric as a~consequence of symmetry of
$\tilde{\matE}^e_{kl}, \matE_l^e$ and $\bigl(\matM^e\bigr)^{-1}$. The products $\boxed{2a}$ and
$\boxed{2b}$ are transposes of each other, hence their sum is again symmetric. The modifications to
the symmetric discrete Laplacian therefore preserve symmetry of the discrete weak form, meaning that
efficient iterative solvers such as the conjugate gradient method can be used to obtain solutions.
%
%
\subsection{Convergence rates comparison: weak vs. strong boundary conditions}
\begin{figure}
  \centering
  \subfloat[][Convergence to exact solution, strong vs. weak boundary conditions on unstructured triangular mesh.]
  {
     \includegraphics[viewport=20 20 440 330,clip,width=0.45\textwidth]{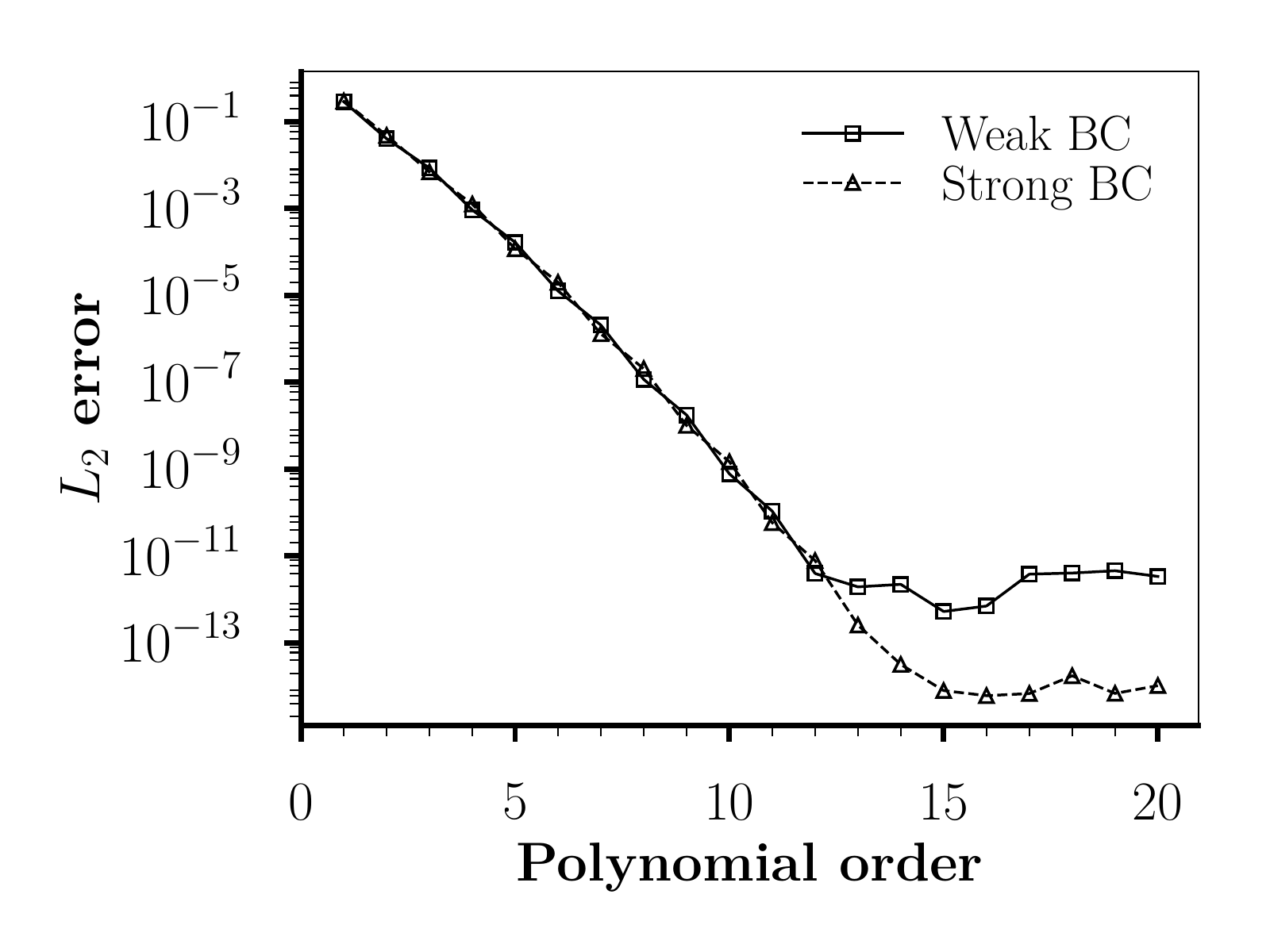}
  }
  \:
  \subfloat[][Convergence to exact solution, strong vs. weak boundary conditions on quadrilateral mesh.]
  {
    \includegraphics[viewport=20 20 440 330,clip,width=0.45\textwidth]{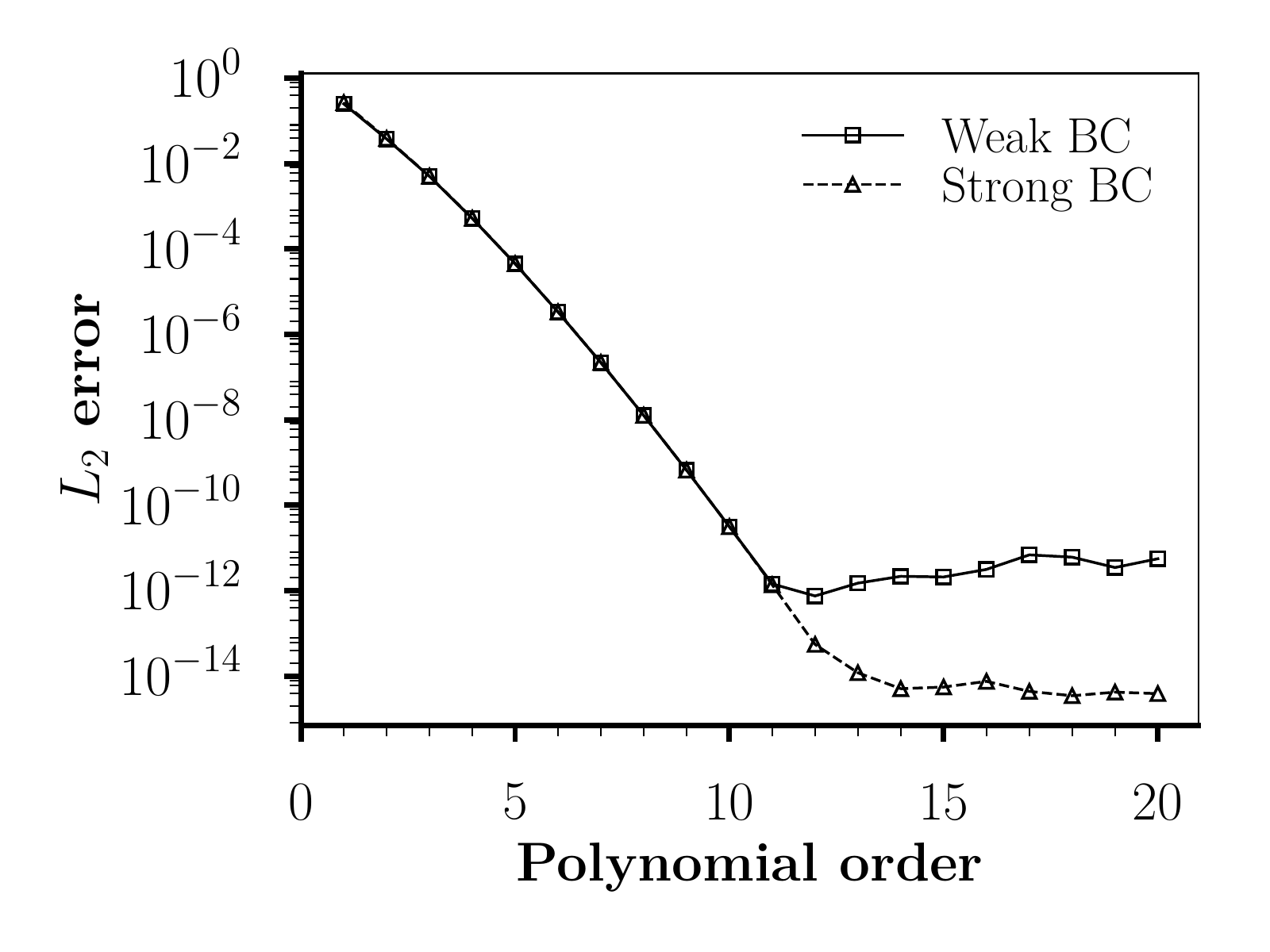}
  }
  \caption{Convergence to the exact solution in $L_2$ norm.}
  \label{fig_convergence_bc}
\end{figure}

%
%
The convergence rates were verified on a~scalar Helmholtz problem with nonzero Dirichlet boundary conditions 
\begin{equation}
  \nabla^2 u - \lambda u = f
\end{equation}
in a~square domain $(-1,1)^2$ with $\lambda = 1$ and $f(x,y)$ chosen so that the exact solution is
of the form
\begin{equation}
  u(x,y) = \sin(10\pi x) \cos(10\pi y) + x + y
\end{equation}
Figure~\ref{fig_convergence_bc} compares the $L^2$ error for polynomial orders varying between 1 and
20 when the Dirichlet boundary conditions are imposed strongly and weakly. The behaviour of both
strong and weak methods produces nearly identical errors up to $p = 12$ on the structured grid and
$p=11$ on triangles. With further increase of polynomial degree of the basis, however, the weak
errors fail to further decrease. The observed differences are not surprising, because the
\HDG{}-based algorithm only penalizes the solution in order to satisfy boundary conditions, while
the strong implementation completely eliminates known degrees of freedom and moves them to the
right-hand side of the linear system, thus fulfilling the boundary conditions exactly by construction.
Furthermore, the stiffness matrix with weak constraints is larger, hence less favourably conditioned
and round-off errors in the linear solver become important as the error values approach the limits
of finite-precision arithmetic on given machine.
%
%
%
\section{Expected Performance}
%
%
\subsection{Cost in terms of FLOPs}
We assume that the discrete Poisson problem is solved in two stages, both of which will significantly 
contribute to the overall CPU time spent in the solver. The stages are:
\begin{enumerate}
  \item \textbf{Assembly and solution of statically condensed system.} This step involves processing 
  unknowns on entity boundaries,  where 'entity' would be each single element in the context of 
  continuous and hybrid discontinuous Galerkin methods and one mesh patch (a group of elements spanned 
  by a~continuous polynomial basis) in the combined continuous-discontinuous Galerkin method. 
  
  The main difference between \CG{} and \HDG{} is that in the continuous case, trace variables are 
  identical to variables located on element boundaries and are shared by neighbouring elements. The 
  \HDG{} method, on the other hand, introduces an additional hybrid  variable, thus requiring more 
  memory storage. This variable is not globally continuous, hence degrees of freedom on face 
  boundaries are duplicated. As a~consequence, the \HDG{} interior solve on each element has to 
  process a~slightly larger local system.
  \item \textbf{Interior solve.} Given solution on entity boundary, the solution in entity interior 
  is reconstructed during this stage. Interior solve involves the inverse of a~potentially large matrix.
\end{enumerate}
Setup costs (for example precomputing and storing the matrix inverses needed in interior solve above) 
are not taken into account.
%
%
\subsection*{Domain Description}
\begin{figure}[htpb]
  \centering
  \includegraphics[width=0.95\textwidth]{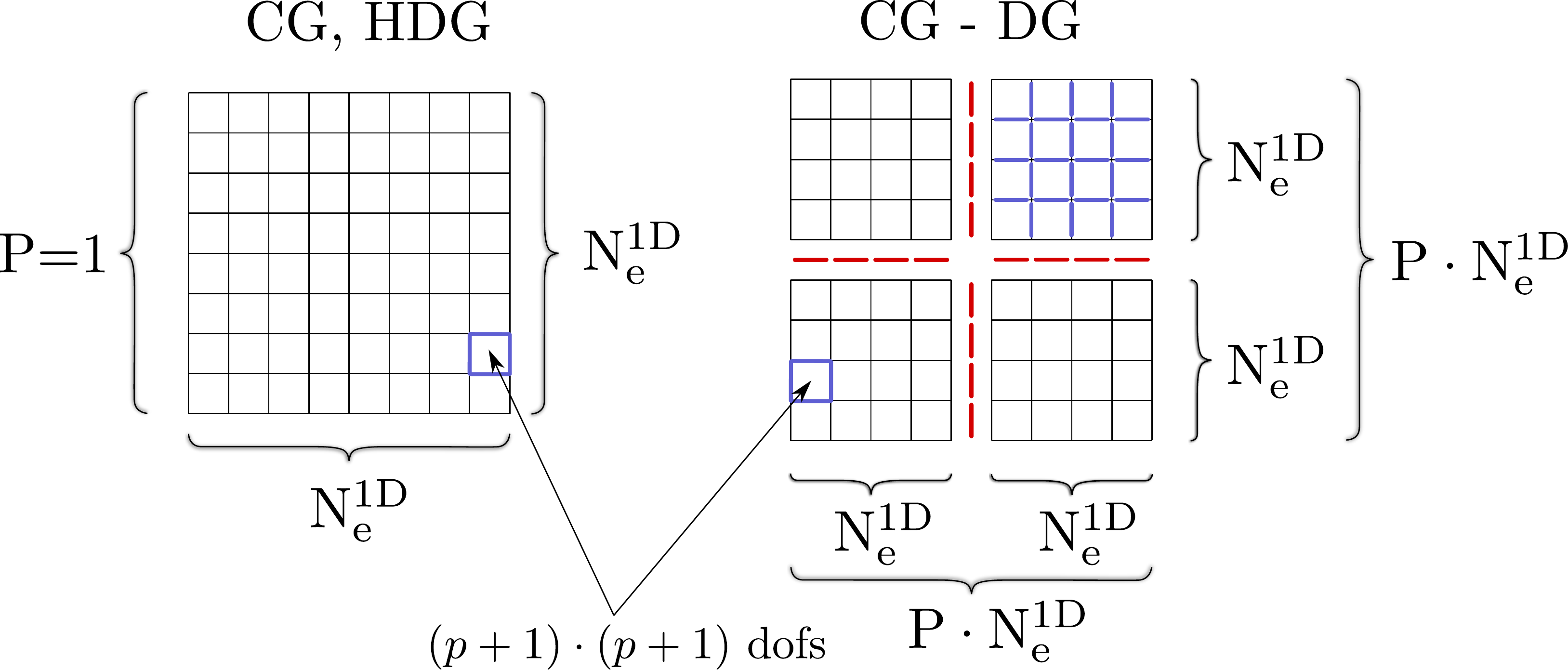}
  \caption{Idealized mesh divided into $P\times P$ patches, each patch containing 
  $\numElemOneD{} \times \numElemOneD{}$ elements of order $p$.}
  \label{fig_dof_counts_cg_dg}
\end{figure}

\begin{figure}[htpb]
  \centering
  \includegraphics[width=0.8\textwidth]{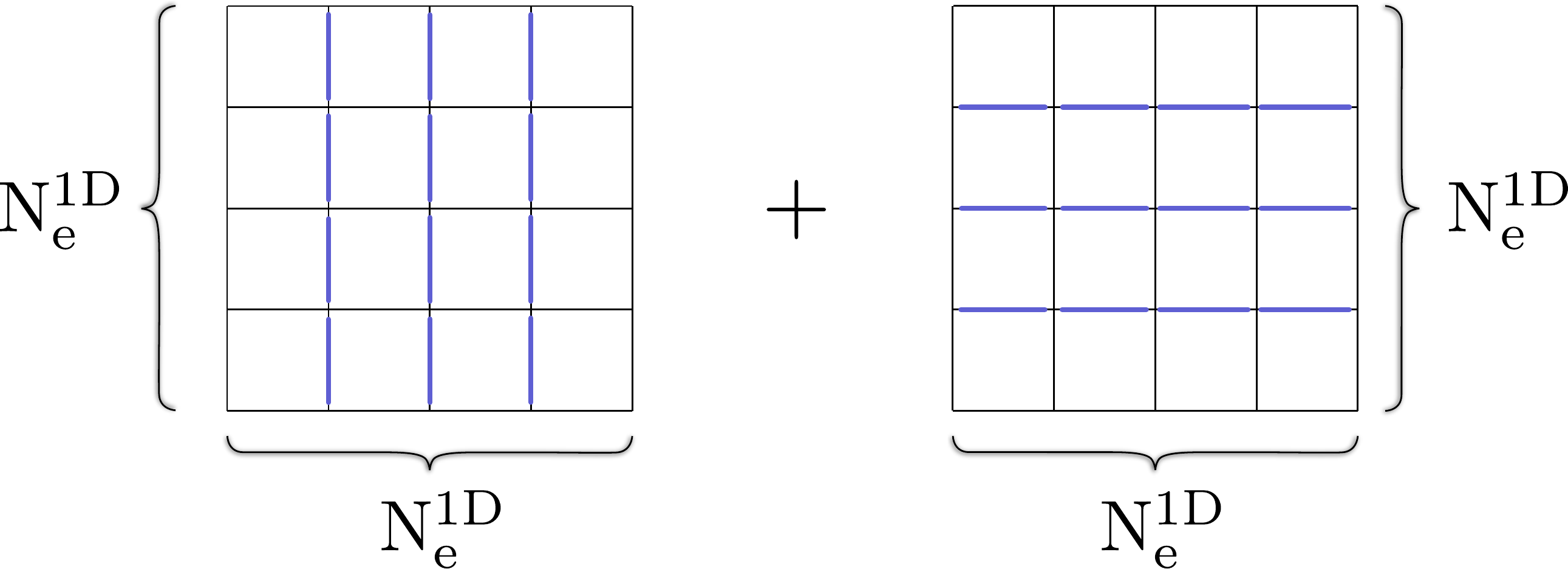}
  \caption{Interior edges (blue) within a patch.}
  \label{fig_edge_count_hdg}
\end{figure}

%
%
%
We assume a~structured grid divided into $P\times P$ patches, each patch consisting of 
$\numElemOneD{} \times \numElemOneD{}$ elements (figure \ref{fig_dof_counts_cg_dg}). Each element 
has a~polynomial basis of degree $p$, i.e. $(p+1)\times (p+1)$ degrees of freedom.
These may or may not be shared with neighbouring elements, depending on the setup (\CG{} vs. \DG{} 
vs. \HDG{}) and global continuity of the polynomial bases. The number of inter-patch edges (red edges 
in figure \ref{fig_dof_counts_cg_dg}) is 
\begin{equation*}
  N_{interpatch}^{edges} = 2 P \cdot (P-1) \cdot \numElemOneD{}, 
\end{equation*}
and each patch contains $N_{patch}^{edges}$ interior edges, with 
\begin{equation*}
  N_{patch}^{edges} = 2 \numElemOneD{} \cdot (\numElemOneD{}-1),
\end{equation*}
see figure \ref{fig_edge_count_hdg}.
\FloatBarrier
%
%
%
%
\subsection*{Stage I: Solution of Statically Condensed System}
\subsubsection*{Continuous Galerkin}
%
Since the total number of elements along each side of the mesh is $P \cdot \numElemOneD{}$ in 2D, the 
total number of unknowns before static condensation (assuming Dirichlet boundary condition everywhere) 
is 
\begin{equation*}
  N^{dof}_{CG} = \bigl(P \cdot \numElemOneD{} \cdot p - 1\bigr)^2.
\end{equation*}
This is also the rank of global system matrix. In case of one-level static condensation, the global 
system  has the form
\begin{equation*}
  \left[\begin{array}{cc} \matM{}_b & \matM{}_c\\ \matM{}_c^{T} & \matM{}_i\end{array}\right]
  \left[\begin{array}{c} \vecx{}_b \\ \vecx{}_i \end{array}\right] = 
  \left[\begin{array}{c} \vecf{}_b \\ \vecf{}_i \end{array}\right]
\end{equation*}
and the rank of $\matM{}_b$ is \emph{approximately} (counting the boundary modes on the skeleton of 
the mesh) equal to
\begin{align*}
  N^{\lambda}_{CG} = (N^{edges}_{interpatch} + P \cdot N^{edges}_{patch})\cdot p 
                  &= (2P\cdot (P-1) \cdot \numElemOneD{} + P \cdot 2\numElemOneD{} \cdot (\numElemOneD{}-1)) \cdot p\\
                  &= 2P\numElemOneD{} (P + \numElemOneD{} - 2)p.
\end{align*}
The remaining values of $u$ in element-interior degrees of freedom can be obtained by inverting 
$(P \cdot \numElemOneD{})^2$ local matrices of rank $p-1$. This means that the total cost of solving 
the \CG{} problem is
\begin{equation*}
  C_{CG} = \mathcal{O}\bigl(cgsolve(P \numElemOneD{}(P + \numElemOneD{}) p)\bigr) + (P\numElemOneD{})^2\cdot \mathcal{O}\bigl((p-1)^3\bigr),
\end{equation*}
where $cgsolve(n)$ is the cost function of solving a~sparse system of rank $n$ with conjugate 
gradients. The cost of the second term is small if the blocks of $\matM_{i}$ are inverted and stored 
during setup phase. The second term in the estimate assumes that the inverse of each diagonal block 
of $\matM{}_i$ costs as much as Gauss elimination/LU decomposition of a~matrix of rank $p-1$, 
which has cubic time complexity.
%
%
%
\subsubsection*{HDG}
%
%
%
The discrete transmission condition (\ref{eq_transmission_condition}) generates a~sparse system of rank
\begin{align*}
  N^{\lambda}_{HDG} = (N^{edges}_{interpatch} + P \cdot N^{edges}_{patch})\cdot (p+1) 
                  &= (2P (P-1) \numElemOneD{} + P\cdot 2\numElemOneD{} (\numElemOneD{}-1)) \cdot (p+1)\\
                  &= 2P\numElemOneD{} (P + \numElemOneD{} - 2)(p+1).
\end{align*}
In addition, we need to invert $(P\numElemOneD{})^2$ local systems $\in \mathbb{R}^{(p+1)\times(p+1)}$ 
as in the \CG{} case. The backsolve is more expensive however, because we have $d$ mixed variables 
$q_1, \ldots q_d$ in $d$ dimensions. The element local inversion can be again precomputed and stored 
during setup. 

The overall cost of solving for all unknowns scales as
\begin{equation*}
  C_{HDG} = \mathcal{O}\bigl(cgsolve(P \numElemOneD{}(P + \numElemOneD{}) (p+1))\bigr) + (P\numElemOneD{})^2\cdot \mathcal{O}\bigl((p+1)^3\bigr).
\end{equation*}
%
%
%
\subsubsection*{Combined CG-DG Solver}
%
%
The number of hybrid degrees of freedom on interfaces between patches is
\begin{equation*}
  N^{\lambda}_{CG-DG} = N^{edges}_{interpatch} (p+1) = 2P (P-1) \numElemOneD{} (p+1).
\end{equation*}
Each patch contains \emph{approximately} $(\numElemOneD{} p)^2$ interior degrees of freedom, hence the total cost is
\begin{equation*}
  C_{CG-DG} = \mathcal{O}\bigl(cgsolve(P^2 \numElemOneD{} (p+1))\bigr) + P^2 \cdot \mathcal{O}\bigl((\numElemOneD{} p)^3\bigr).
\end{equation*}
In the limiting case where each patch coincides with one single element (i.e. $P:= \numElemOneD{}$ and $\numElemOneD{} = 1$), the three estimates 
$C_{CG}$, $C_{HDG}$ and $C_{CG-DG}$ predict the same asymptotic cost.
%
%
%
\subsubsection*{Cost of Solving the Statically Condensed System}
%
%
\textsf{Standard HDG Algorithm}
\vspace*{1em}
\newline
The cost of linear solve in the \PCG{} (preconditioned conjugate gradient) solver will mainly depend 
on the cost of evaluating matrix-vector multiplications. For a~matrix of rank $n$, this cost is 
$\mathcal{O}(n^2)$. Nektar++ solves the statically condensed system in matrix-free manner by performing 
the above matrix-vector multiplications \emph{element-wise} and then summing them together. Suppose 
the (structured) mesh consists of quadrilaterals in 2D and hexahedra in 3D. Furthermore, we will 
assume that the triangular mesh is obtained by splitting each quadrilateral into 2 triangles and 
tetrahedral mesh is created by dividing each hexahedron into 6 tetrahedra.

The number of trace degrees of freedom of one element is
\begin{itemize}
  \item $3 \cdot (p+1)$ for \textbf{triangles}
  \item $4 \cdot (p+1)$ for \textbf{quadrilaterals}
  \item $4 \cdot \frac{(p+1)(p+2)}{2}$ for \textbf{tetrahedra}
  \item $6 \cdot (p+1)^2$ for \textbf{hexahedra}
\end{itemize}
Under this assumption, \emph{one matrix-vector multiplication} for the whole system (but performed 
on element-wise basis) will take
\begin{itemize}
  \item $\mathcal{O}\bigl(2(\numElemOneD{})^2\bigl[3 \cdot (p+1)\bigr]^2\bigr) 
              = \mathcal{O}\bigl(18 (\numElemOneD{})^2 (p+1)^2\bigr)$ operations on \textbf{triangles} in 2D
  \item $\mathcal{O}\bigl((\numElemOneD{})^2\bigl[4 \cdot (p+1)\bigr]^2\bigr) 
              = \mathcal{O}\bigl(16 (\numElemOneD{})^2 (p+1)^2\bigr)$ operations on \textbf{quadrilaterals} in 2D
  \item $\mathcal{O}\bigl(6(\numElemOneD{})^3\bigl[2 \cdot (p+1)(p+2)\bigr]^2\bigr) 
              = \mathcal{O}\bigl(24 (\numElemOneD{})^3 (p+1)^2(p+2)^2\bigr)$ operations on \textbf{tetrahedra} in 3D
  \item $\mathcal{O}\bigl((\numElemOneD{})^3\bigl[6 \cdot (p+1)^2\bigr]^2\bigr) 
              = \mathcal{O}\bigl(36 (\numElemOneD{})^3 (p+1)^4\bigr)$ operations on \textbf{hexahedra} in 3D
\end{itemize}
\vspace*{1em}
\textsf{HDG Algorithm Applied to Groups of Continuously Connected Elements}
\vspace*{1em}
\newline
Now suppose that the trace system is built between patches and each patch has $\numElemOneD \times 
\numElemOneD$ quadrilaterals in 2D and $\numElemOneD \times \numElemOneD \times \numElemOneD$ 
hexahedra in 3D. The number of unknowns on the trace of one patch now becomes
\begin{itemize}
  \item $4 \cdot \numElemOneD \cdot p$ in 2D (\textbf{triangles} and \textbf{quadrilaterals}) and
  \item $6 \cdot (\numElemOneD)^2 \cdot p^2$ in 3D (\textbf{tetrahedra} and \textbf{hexahedra}),
\end{itemize}
which will require 
\begin{itemize}
  \item $\mathcal{O}\bigl(16 (\numElemOneD)^2\cdot p^2\bigr)$ operations per matrix-vector multiplication in 2D and
  \item $\color{red}{\mathcal{O}\bigl(36 (\numElemOneD)^4\cdot p^4\bigr)}$ operations in 3D
\end{itemize}
\textbf{This means that the \PCG{} algorithm in \CG{}-\HDG{} case scales one order worse when measured 
in terms of number of elements along patch face \big($\mathcal{O}\bigl((\numElemOneD)^4\bigr)$\big) 
than the standard \HDG{} algorithm \big($\mathcal{O}\bigl((\numElemOneD)^3\bigr)$\big)}.
\begin{remark}
Note that the number on the surface of the patch is the same for triangles and quadrilaterals and for 
tetrahedra and hexahedra, respectively. For a~continuous expansion, the number of DOFs on one 
quadrilateral face of a~hexahedron is $(p+1)^2$, and $2 \cdot \frac{(p+1)(p+2)}{2} - (p+1) = (p+1)^2$ 
for two triangles covering the same quadrilateral face. 
\end{remark}
\begin{figure}[htpb]
  \centering
  \includegraphics[width=0.49\textwidth]{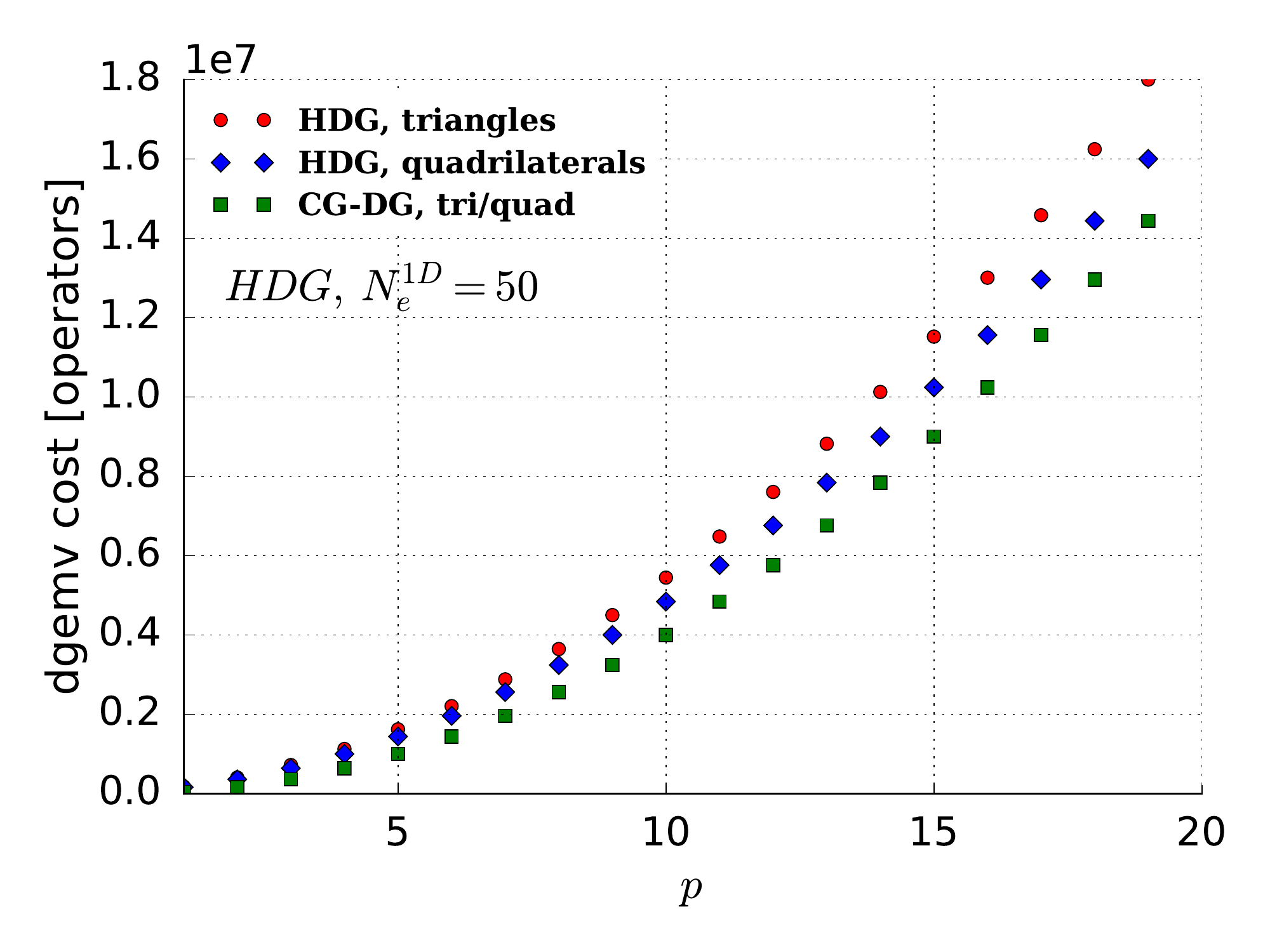}
  \includegraphics[width=0.49\textwidth]{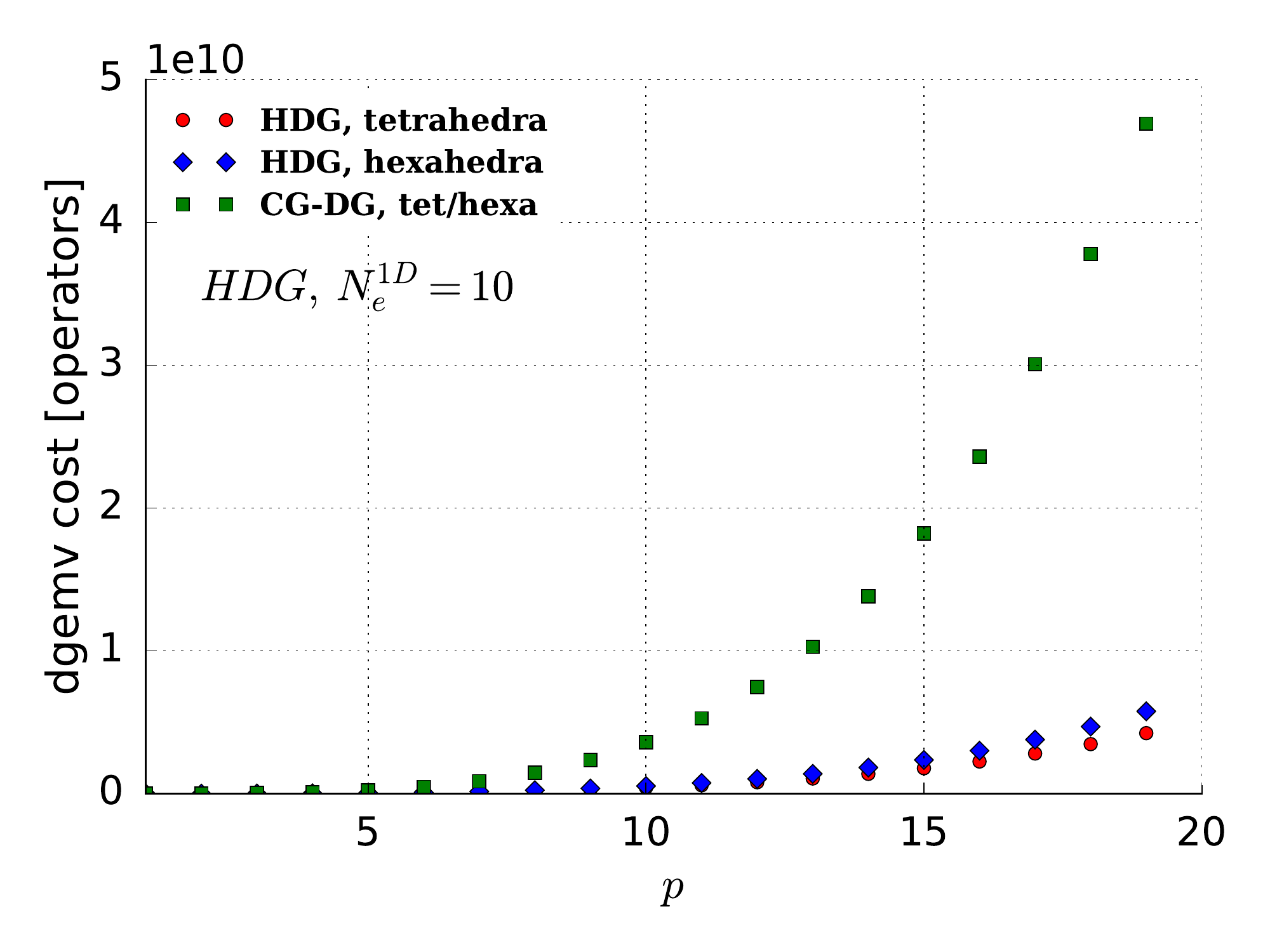}
  \caption{Asymptotic cost of matrix-vector multiplication measured by operation counts for \HDG{} and 
  combined \CG{}-\HDG{} methods.}
  \label{fig_cost_dgemv_2D_3D}
\end{figure}
%
%
%
\subsection*{Stage II: Interior Solve}
%
%
The reconstruction of interior degrees of freedom involves the solution of a~linear system with the 
matrix
\begin{equation*}
  \matA^e = \left(\begin{array}{ccc} 
                 \sum\limits_{l=1}^{N_b^e} \tau^{(e,l)} \matE_l^e & -\matD_1^e & -\matD_2^e\\
                 \bigl(\matD_1^e\bigr)^T & \matM^e & \vecZero\\
                 \bigl(\matD_2^e\bigr)^T & \vecZero & \matM^e
            \end{array}\right)
\end{equation*}
The superscript $^e$ no longer refers to a~single element as was the case of \HDG{}. For \CG{}-\HDG{} 
method, all the blocks in $\matA$ are a~result of a~continuous Galerkin discretization in the whole 
partition/patch. The sparse matrix $\matA^e$ is potentially large and its explicit inverse will be 
dense, i.e. require significant storage.
\newline\\
\textbf{The expensive interior solve together with increased operation count when inverting the 
statically condensed system in 3D indicates that the benefit of reduced communication pattern in 
continuous-discontinuous discretization might be outweighed by extra CPU cost and there is relatively 
little performance (if any) to be gained by combining the continuous and discontinuous Galerkin 
discretization into one hybrid solver.}
\subsection{Cost in terms of memory requirements}
%
%
%
\subsection*{Stage I: Solution of Statically Condensed System}
%
We again assume that the global system is solved by performing multiple \PCG{} iterations, where 
global matrix-vector multiply is executed in matrix-free fashion. For each elemental multiplication, 
the data containing the input matrix and vector and resulting vector must be loaded into processor 
cache. We discuss amount of data transferred for each method \emph{during one \PCG{} iteration} here.
\subsubsection*{Continuous Galerkin}
On triangles, one element contains $3\cdot p$ trace degrees of freedom (this is irrespective of 
global continuity of the solution, because we perform the multiplication element-by element and hence 
whether the DOFs are shared with neighbouring elements or not is irrelevant). The elemental statically 
condensed matrix has rank $(3 \cdot p)^2$ and the amount of data to move in and out of cache is 
therefore
\begin{equation*}
 (3 \cdot p)^2  + 2 \cdot (3 \cdot p) = 9p^2  + 6p
\end{equation*}
(The term $2\cdot(3\cdot p)$ takes into account two vectors needed for elemental matrix-vector 
multiplication.) This is repeated for each element, and therefore the total number of floating-point 
values transferred is
\begin{equation*}
  (2\cdot \numElemOneD) (9p^2  + 6p)
\end{equation*}
Repeating similar calculation for other element types, we arrive to the following estimates for 
continuous Galerkin method and different element types:
\begin{itemize}
  \item \textbf{Triangles}: $N_{CG, tri} = (2(\numElemOneD{})^2 (9p^2 + 6p)$ floating point values
  \item \textbf{Quadrilaterals}: $N_{CG, quad} = (\numElemOneD{})^2 (16p^2 + 8p)$ floating point values
  \item \textbf{Tetrahedra:} $N_{CG, tet} = (6\cdot \numElemOneD{})^3 \bigl((2p(p+1))^2 + 4p(p+1)\bigr)$ floating point values
  \item \textbf{Quadrilaterals:} $N_{CG, hex} = (\numElemOneD{})^3 \bigl((6p^2)^2) + 12p^2)\bigr)$ floating point values
\end{itemize}
\subsubsection{Discontinuous Galerkin}
Data for \HDG{} are very similar with the exception that the trace values are discontinuous, which 
means that the elemental matrices are slightly bigger:
\begin{itemize}
  \item \textbf{Triangles:} $N_{HDG, tri} = (2(\numElemOneD{})^2 (9(p+1)^2 + 6(p+1))$ floating point values
  \item \textbf{Quadrilaterals:} $N_{HDG, quad} = (\numElemOneD{})^2 (16(p+1)^2 + 8(p+1))$ floating point values
  \item \textbf{Tetrahedra:} $N_{HDG, tet} = (6\cdot \numElemOneD{})^3 \bigl((2(p+1)(p+2))^2 + 4(p+1)(p+2)\bigr)$ floating point values
  \item \textbf{Quadrilaterals:} $N_{HDG, hex} = (\numElemOneD{})^3 \bigl((6(p+1)^2)^2 + 12(p+1)^2\bigr)$ floating point values
\end{itemize}
\subsubsection{CG-HDG}
The hybrid \CG{}-\HDG{} method has a~system matrix of rank $(4\cdot \numElemOneD{} \cdot p)$, which 
means that the number of floating point values involved in one matrix-vector multiply will be\\

\vspace*{2mm}
$N_{CG-HDG, 2D} = (4\cdot \numElemOneD{} \cdot p)^2 + (8\cdot \numElemOneD{} \cdot p) 
                = 16(\numElemOneD{})^2 p^2 + 8 \numElemOneD{} p$ in 2D\\

and similarly in 3D, where the number of DOFs on patch surface is $6 \cdot (\numElemOneD{})^2 \cdot p^2$\: :\\

\vspace*{2mm}
$N_{CG-HDG, 3D} = (6\cdot (\numElemOneD{})^2 \cdot p^2)^2 + (12\cdot (\numElemOneD{})^2 \cdot p^2)$ in 3D.
%
%
%
\subsection*{Stage II: Interior Solve}
%
%
\subsubsection*{Continuous Galerkin}
The number of interior degrees of freedom in one high-order triangle is $(p+1)(p+2)/2 - 3p = (p-2)(p-1)/2$ 
and we suppose that this is the rank of elemental Schur complement which has to be inverted and 
stored. In the case of continuous Galerkin system, the element-interior matrix, left- and right- hand 
side vectors hold 
\begin{equation*}
 N_{CG, tri} = ((p-2)(p-1)/2)^2 + 2 \cdot \bigl((p-2)(p-1)/2\bigr)
\end{equation*}
This cost has to be multiplied by number of elements present in the mesh.
Cost for different element shapes is summarized below
\begin{itemize}
  \item \textbf{Triangles}: $N_{CG, tri} = (2(\numElemOneD{})^2 \bigl[((p-2)(p-1)/2)^2 + (p-2)(p-1)\bigr]$ floating point values
  \item \textbf{Quadrilaterals}: $N_{CG, quad} = (\numElemOneD{})^2 \bigl[(p-1)^2 + 2(p-1)\bigr]$ floating point values
%
%
\end{itemize}
\subsubsection*{Discontinuous Galerkin}
In \HDG{}, each elements has its 'own' DOFS not shared with the hybrid variable, hence elemental 
matrices are again slightly bigger:
\begin{itemize}
  \item \textbf{Triangles}: $N_{HDG, tri} = (2(\numElemOneD{})^2 \bigl[((p+1)(p+2)/2)^2 + (p+1)(p+2)\bigr]$ floating point values
  \item \textbf{Quadrilaterals}: $N_{HDG, quad} = (\numElemOneD{})^2 \bigl[(p+1)^2 + 2(p+1)\bigr]$ floating point values
%
%
\end{itemize}
\subsubsection*{CG-HDG}
The 'interior matrix' is sparse, but involves all DOFS of the patch, whose count is approximately 
$(\numElemOneD{})^2 p^2$. The matrix and corresponding storage would then be
\begin{equation*}
  \color{red}{N_{CG-HDG, quad} = \bigl((\numElemOneD{})^2 p^2\bigr)^2 + 2 (\numElemOneD{})^2 p^2}
\end{equation*}
Which is again orders of magnitude worse estimate than for \CG{} and \HDG{}.
%
%
%
\section{Conclusion}
\label{sec_conclusion}
This paper proposes a~new method for combining the \CG{} and \HDG{} solvers and 
derives an algorithm for the imposition of Dirichlet boundary conditions for elliptic 
PDEs of Helmholtz type which enforces the constraints weakly, i.e. by amending the underlying weak 
form with penalty terms instead of lifting known boundary values from the linear system.

The presented technique is conceptually based on hybrid Discontinuous Galerkin method, but replaces 
the polynomial space typically used in element interiors (a~finite element basis defined in single 
element) by a~piecewise continuous multi-element Galerkin expansion. We demonstrate that the method 
is conceptually feasible and it combines some attractive features of \CG{} and \HDG{}, but it fails 
to deliver the expected performance. Even if we stored the inverted local solvers to effectively recover 
degrees of freedom located on each mesh partition, the cost of assembly and solution of the 
discrete transmission condition in three dimensions remains prohibitively expensive.
%
\section*{Acknowledgments}
%
%
Martin Vymazal was supported by a European Commission Horizon 2020 project grant entitled “ExaFLOW: Enabling Exascale Fluid Dynamics Simulations” (grant reference 671571).
%
%
%
%

%
\end{document}